\documentstyle[12pt,a4]{article}

\font\mybb=msbm10 at 12pt
\def\bb#1{\hbox{\mybb#1}}
\def\Z {\bb{Z}}

\newcommand{\VEV}[1]{\left\langle #1 \right\rangle}
\newcommand{\vs}{\vspace*}
\newcommand{\hs}{\hspace*}
\newcommand{\wt}{\widetilde}
\newcommand{\del}{\partial}
\newcommand{\ra}{\rightarrow}

\newcommand{\sq}{\sqrt{2}\,}
\newcommand{\nn}{\nonumber}

\newcommand{\mybar}[1]%
        {\kern 0.8pt\overline{\kern -0.8pt#1\kern -0.8pt}\kern 0.8pt}
\newcommand{\drawsquare}[2]{\hbox{%
\rule{#2pt}{#1pt}\hskip-#2pt
\rule{#1pt}{#2pt}\hskip-#1pt
\rule[#1pt]{#1pt}{#2pt}}\rule[#1pt]{#2pt}{#2pt}\hskip-#2pt
\rule{#2pt}{#1pt}}

\newcommand{\Yfund}{\raisebox{-.5pt}{\drawsquare{6.5}{0.4}}}
\newcommand{\Ybarfund}{\mybar{\raisebox{-.5pt}{\drawsquare{6.5}{0.4}}}}

\newcommand{\Yasymm}{\raisebox{-3.5pt}{\drawsquare{6.5}{0.4}}\hskip-6.9pt%
        \raisebox{3pt}{\drawsquare{6.5}{0.4}}}
\newcommand{\Ybarasymm}{\mybar{\raisebox{-3.5pt}{\drawsquare{6.5}{0.4}}
    \hskip-6.9pt\raisebox{3pt}{\drawsquare{6.5}{0.4}}}}

\begin{document}
\begin{titlepage}

\begin{flushright}
KUNS-1511\\
HE(TH)98/09\\
hep-ph/9805457\\
May, 1998
\end{flushright}

\vs{6ex}
\begin{center}
{\large \bf
A Model with Simultaneous Dynamical Breaking of Supersymmetry and GUT
Gauge Symmetry
}
\vspace{12ex}

Takayuki Hirayama
\footnote{e-mail: hirayama@gauge.scphys.kyoto-u.ac.jp},
Naoya Ishimura
\footnote{e-mail: ishimura@gauge.scphys.kyoto-u.ac.jp}
and
Nobuhiro Maekawa
\footnote{e-mail: maekawa@gauge.scphys.kyoto-u.ac.jp}
\vspace{0.5cm}

{\it Department of Physics, Kyoto University,\\
     Kyoto 606-8502, Japan}
\end{center}
\vspace{1cm}

\vs{6ex}
\begin{abstract}
We try to construct a model in which supersymmetry and grand 
unified gauge symmetry are dynamically broken at the same time. In
this model SUSY breaking is mediated mainly by massive vector
multiplet, and a new solution for the $\mu$ problem is proposed.
\end{abstract}


\end{titlepage}


If supersymmetry (SUSY) is a real solution for gauge hierarchy
problem, we should understand how SUSY is broken and why the SUSY
Higgs mass is around the SUSY breaking scale, which is much less than
the grand unified scale or the Planck scale. One of the most
interesting scenarios of understanding the SUSY breaking is gauge
mediated SUSY breaking scenario \cite{dine}. In the last few years,
using the recent development on SUSY gauge theories \cite{seiberg,izawa},
various dynamical SUSY breaking models are examined
\cite{dine,murayama-model,gmsb,murayama,giudice2}.
Unfortunately, in the gauge mediated SUSY breaking scenario, we have
not understood well why the SUSY Higgs term is around the SUSY
breaking scale, though various attempts are considered
\cite{dine,murayama,mu}. One possible solution for the $\mu$ problem
is to introduce the singlet field which is coupled with the Higgs
fields, and to realize that the vacuum expectation value (VEV) which
becomes the SUSY Higgs mass, is determined by the SUSY breaking
parameters \cite{dine}. However, it is dangerous because the radiative
correction generally breaks the gauge hierarchy when the singlet field
couples to the heavy fields \cite{singlet}.

On the point of SUSY grand unified theories (GUT) we should understand
not only the mechanism of the doublet-triplet splitting \cite{dt} but
the consistency between GUT formalism and the solution of the $\mu$
problem. For example, the singlet field introduced for the $\mu$
problem is also coupled with the colored Higgs, which breaks gauge
hierarchy.

In this paper, we examine a model in which SUSY and GUT gauge
symmetry are simultaneously broken and the $\mu$ problem can be
solved. To break SUSY and GUT gauge symmetry at the same time,
it is natural that SUSY breaking
is mediated by the massive vector multiplet (gauge messenger)
\cite{giudice}.

The gauge messenger can work only when the F-term of
the Higgs, which breaks a gauge symmetry, is non-zero, namely when the
spectrum of the massive gauge multiplets deviates from the SUSY limit.
Since the original gauge symmetry should include the standard gauge
symmetry for inducing all needed soft SUSY breaking parameters, it is
natural that GUT gauge symmetry is adopted as the original one.

We examine a following model with the gauge symmetry $SU(5)_G\times
SU(3)_H\times SU(2)_H \times U(1)_H$
in addition to the strong gauge group $Sp(8)$
\footnote{The rank of $Sp(2N)$ is $N$.}.
The quantum numbers of introduced fields are
\begin{center}
$
\begin{array}{|c||ccc|cccc|}
  \hline
  &Sp(8) &SU(5)_G   &SU(5)_H       &U(1)_V&U(1)_A&U(1)_R&\Z_2 \\
  \hline
  \wt Q   &\Yfund&\Ybarfund        &1         &1 &0 &0   &0 \\
  Q       &\Yfund&1                &\Yfund    &-1&0 &0   &0 \\
  X       &1     &\Yasymm          &1         &-2&0 &2   &0 \\
  \Sigma  &1     &\Yfund           &\Ybarfund &0 &0 &2   &0 \\
  \wt X   &1     &1                &\Ybarasymm&2 &0 &2   &0 \\        
  \wt 5_i &1     &\Ybarfund\times 3&1         &0 &3/2&-1 &1 \\
  10_i    &1     &\Yasymm\times 3  &1         &0 &-1/2 &1&0 \\
  H_a     &1     &\Yfund\times 2   &1         &0 &1 &0   &0 \\
  \phi&1     &\Yfund        &1         &0 &1 &0  &1 \\
  \wt \phi&1     &\Ybarfund        &1         &0 &-1 &2  &1 \\
  \wt H_1 &1     &1         &\Ybarfund&0 &-1 &2  &\left\{
    \begin{array}{cc}
      \wt T_1&0\\
      \wt D_1&1
    \end{array}\right.\\
  \wt H_2 &1     &1         &\Ybarfund&0 &-1 &2  &\left\{
    \begin{array}{cc}
      \wt T_2&0\\
      \wt D_2&0
    \end{array}\right.\\
  \hline
\end{array}
$  
\end{center}
where $\wt T$ and $\wt D$ are triplet and doublet superfields,
respectively. We write $SU(3)_H\times SU(2)_H\times U(1)_H$ as
$SU(5)_H$ in short
\footnote{
Though the quantum numbers of the SUSY breaking sector are almost same
as in Murayama's model \cite{murayama-model}, the dynamical assumption
is different.
}.
The index $i=1,2,3$ is the generation number and the index $a$ is 
1 or 2. Here $U(1)_R$ is R-symmetry and the global symmetry
$U(1)_V\times U(1)_A\times U(1)_R\times \Z_2$ has no anomaly under the
strong $Sp(8)$ gauge symmetry.

Although we adopt a non simple group for the GUT group, the ordinary
matter is unified in the $SU(5)_G$ gauge group. 
The generic superpotential with the canonical dimension less than 6 is
\begin{eqnarray}
  W&=& 
  \lambda (Q\wt Q) \Sigma +\kappa (QQ) \wt X
  +\wt \kappa (\wt Q\wt Q) X \nn\\
  &&+M_\phi\phi\wt \phi +\frac{1}{M}(\wt QQ)(a_{ab}H_a\wt T_b 
  +b_aH_a\wt D_2+c\phi \wt D_1) \nn\\
  &&+(y_u)_{ij}^a 10_i10_jH_a +(y_d)_{ij}10_i\wt 5_j\wt \phi 
  +(\wt y_d)_{ij}\frac{1}{M^2}(\wt QQ)10_i\wt 5_j\wt D_1 ~,
 \label{0}
\end{eqnarray}
where $M_{\phi}$ is the same or smaller than a scale of new physics
$M$. Here we shorten the description of the superpotential
\footnote{In this superpotential, there is a $U(1)_{PQ}$ symmetry
  instead of $\Z_2$ symmetry. By adding some terms later in order to
  make more realistic model, only $\Z_2$ symmetry remains.
  }, for example, $\lambda (Q\wt Q)\Sigma$ is divided to the $SU(3)_H$
part $\lambda_1 (Q\wt Q)_3\Sigma_3$ and the $SU(2)_H$ part $\lambda_2
(Q\wt Q)_2\Sigma_2$. When the coupling of $Sp(8)$ gauge group becomes
large at the scale $\Lambda$, where $SU(5)_G \times SU(3)_H\times
SU(2)_H\times U(1)_H$ is weak, $Sp(8)$ gauge theory is known to
confine. Therefore the effective superpotential under the scale
$\Lambda$ can be written in terms of the composite fields
\begin{eqnarray}
  \Sigma_Q \sim \frac{\wt QQ}{\Lambda} \hs{3ex} 
  X_Q\sim \frac{QQ}{\Lambda} \hs{3ex}
  \wt X_Q\sim \frac{\wt Q\wt Q}{\Lambda} ~,\label{3}
\end{eqnarray}
as
\begin{eqnarray}
  W_{eff}&=& \frac{Y}{\Lambda^3}\left( \Sigma_Q^5 
    +X_Q\Sigma_Q^3\wt X_Q
    +X_Q^2\Sigma_Q\wt X_Q^2 -\Lambda^5 \right) \nn \\
  &&+\lambda \Lambda\Sigma_Q \Sigma 
  +\kappa \Lambda X_Q \wt X
  +\wt \kappa \Lambda \wt X_Q X \nn\\
  &&+M_\phi\phi\wt \phi +\frac{\Lambda}{M}\Sigma_Q(a_{ab}H_a\wt T_b 
  +b_aH_a\wt D_2+c\phi \wt D_1) \nn\\
  &&+(y_u)_{ij}^a 10_i10_jH_a +(y_d)_{ij}10_i\wt 5_j\wt \phi 
  +(\wt y_d)_{ij}\frac{\Lambda}{M^2}\Sigma_Q10_i\wt 5_j\wt D_1 ~,
  \label{1}
\end{eqnarray}
where $Y$ is an auxiliary field. Then we can see that SUSY is
dynamically broken from the first and second lines in eq.(\ref{1}).
Under the constraint
\footnote{
  We use the same symbol for the superfield and its scalar component.
}
\begin{equation}
  \Sigma_Q^5 +X_Q\Sigma_Q^3\wt X_Q
    +X_Q^2\Sigma_Q\wt X_Q^2 -\Lambda^5 =0 ~,\label{const}
\end{equation}
the F-term equations
\begin{eqnarray}
    \frac{\del W}{\del \Sigma} &=& \lambda\Lambda\Sigma_Q =0 ~,\\
    \frac{\del W}{\del X} &=& \kappa\Lambda \wt X_Q =0~,\\
    \frac{\del W}{\del \wt X} &=& \wt \kappa\Lambda X_Q =0~,
\end{eqnarray}
cannot be satisfied simultaneously, i.e., SUSY is spontaneously
broken \cite{izawa}. If we take $\lambda \ll \kappa,\wt \kappa$, 
then it is reasonable to take the vacuum 
\begin{eqnarray}
 &\VEV{\Sigma_Q}\sim\Lambda  ~\ra ~\VEV{F_{\Sigma}}
 \sim \lambda\Lambda^2  \neq 0 ~,\label{4}\\
 &\VEV{X_Q}=\VEV{\wt X_Q}=0 ~.
\end{eqnarray}
Since we have no information on the K\"ahler potential of $\Sigma_Q$,
we only assume that the potential has a local minimum point
\begin{equation}
\VEV{\Sigma_Q}=\left(\begin{array}{ccccc}
                    v_1 &   0 &   0 &   0 &   0 \\
                    0   & v_1 &  0  &  0  &  0  \\
                    0   &  0  & v_1 &  0  &  0  \\
                     0  &  0  &  0  & v_2 &  0  \\
                     0  &  0  &  0  &  0  & v_2 \\
                     \end{array}\right),\quad
\VEV{\Sigma}=\left(\begin{array}{ccccc}
                    \wt v_1 &   0 &   0 &   0 &   0 \\
                    0   &\wt v_1 &  0  &  0  &  0  \\
                    0   &  0  &\wt v_1 &  0  &  0  \\
                     0  &  0  &  0  &\wt v_2 &  0  \\
                     0  &  0  &  0  &  0  &\wt v_2 \\
                     \end{array}\right),\label{5}
\end{equation}
under the constraints that all F-terms equal zero except $F_{\Sigma}$
\footnote{
  Since this conditions require $\wt v_1=\wt v_2$, D-flatness is not
  satisfied in this model. We have various solutions for this
  problem. For example, we introduce a singlet $S$ and vector like
  fields $\wt N$, $N$ which has the same quantum number of
  $\Sigma_2$. And we add $SN(Q\wt Q)_2$ and $N\wt N$ to the superpotential.
}. Here $v_1$, $v_2$, $\wt v_1$ and $\wt v_2$ are of the order of
$\Lambda$.

The gauge group $SU(5)_G\times SU(3)_H\times SU(2)_H\times
U(1)_H$ is broken to the standard gauge group $SU(3)_C\times SU(2)_L
\times U(1)_Y$ by the above VEVs. One
combination of $\Sigma_Q$ and $\Sigma$ is the Higgs field eaten by the 
broken gauge freedom. (Since the VEV of the $\Sigma$ field also breaks
the $U(1)_R$ symmetry, R-axion also exists. Following the discussion of 
Bagger, Poppitz and Randall \cite{r-axion},
 the R-axion can be massive because of the explicit 
breaking from some unknown mechanism which realizes the smallness of
the cosmological constant.)
It is known that since the F-term of the Higgs is non zero, 
the massive gauge multiplet can mediate the 
SUSY breaking to the ordinary matter. Note that the 
vector-like chiral multiplet does not work in this model at this
stage. From the rough estimation of the SUSY breaking parameter
\begin{eqnarray}
  M_{SB}\sim 10^{-1}\frac{\VEV{F_{\Sigma}}}{\VEV{\Sigma}}\sim 10^2 ~
  {\rm GeV},
\end{eqnarray}
$\lambda \sim 10^{-13}$ is required because the scale $\Lambda$ should 
be taken as the GUT scale. As discussed later the smallness
of the coupling $\lambda$ is naturally realized.

Before the discussing the $\mu$ term, we study how doublet-triplet
splitting can be realized in this model.
The mass matrices of doublet (represented by $D$) and of triplet
(represented by $T$) Higgs are
\begin{eqnarray}
  \raisebox{-3ex}{$\left.\begin{array}{cc}
      +& D_1\\
      +& D_2\\
      -& D_{\phi}
      \end{array}\right( $}
  \begin{array}{ccc}
    -&+&-\\
    \wt D_1 &\wt D_2&\wt D_{\phi}\\
      0& \epsilon^2M & 0\\
     0&\epsilon^2M&0 \\
    \epsilon^2M& 0 & M_{\phi}
  \end{array}
  \raisebox{-3ex}{$ \left)  \begin{array}{c}
        \\
        \\
        \\
      \end{array}\right.$}
  &&
  \raisebox{-3ex}{$\left.\begin{array}{cc}
        +&T_1\\
        +&T_2\\
        -&T_{\phi}
      \end{array}\right( $}
  \begin{array}{ccc}
    +&+&-\\
    \wt T_1 &\wt T_2&\wt T_{\phi}\\
    \epsilon^2M &\epsilon^2M&0\\
    \epsilon^2M&\epsilon^2M&0\\
    0&0&M_{\phi}
  \end{array}
  \raisebox{-3ex}{$ \left) \begin{array}{c}
        \\
        \\
        \\
      \end{array}\right.,$}
\end{eqnarray}
where $\epsilon=\Lambda/M$ and "$+$" and "$-$" are the $\Z_2$ parity.
The rank of the matrices is two for doublet Higgs and three for
triplet Higgs, i.e., doublet-triplet splitting is realized. The light
Higgs are almost
\begin{eqnarray}
  H_u&=&\frac{1}{\sq}(D_1-D_2) ~,\\
  \wt H_d&=&\wt D_1-\epsilon^2\frac{M}{M_\phi}\wt D_{\phi}~.
\end{eqnarray}
Therefore the Yukawa coupling of the Up type quark with the Higgs
field $H$ is $O(1)$, while the Yukawa coupling of the Down type quark
is $O(\epsilon^2\frac{M}{M_\phi})$. Because of the asymptotically
non-free gauge theory, the top Yukawa coupling is running to the value
on the infra-red fixed point $\sim 1.25$, therefore
$\tan \beta =\VEV{H_u}/\VEV{\wt H_d}$ is $\sim 1.15$
\footnote{
  We take the running top quark mass $m_t(m_t)\sim 165$ GeV.
}. The bottom Yukawa
coupling at the low energy scale is about ten times larger than at the
$\Lambda$ scale. As discussed later $\epsilon$ is $10^{-(1\sim 2)}$ and
the hierarchy between the Up and Down type quark Yukawa is realized.
Though it is obvious that the deformation is needed for more realistic
Yukawa coupling, we do not discuss more on the Yukawa coupling.

Next we would like to discuss on the smallness of the coupling
$\lambda$ which determines the SUSY breaking scale and on the
supersymmetric Higgs mass. The Higgs mass should be around the SUSY
breaking scale, though it vanishes in this model by $\Z_2$ symmetry. If
we would like to explain the $\mu$ problem, we should solve the above
two problems at the same time. Is it possible? Yes, in order to explain
the above two problems simultaneously, we change the assignment of the
$\Z_2$ charge;
\begin{center}
  $
  \begin{array}{|c||c|}
    \hline
    &\Z_2 \\
    \hline
    \wt Q   &0\ra 1 \\
    \wt T_1&0\ra 1\\
    \wt D_1&1\ra 0\\
    \wt T_2&0\ra 1\\
    \wt D_2&0\ra 1\\
    \hline
  \end{array}
  $  
\end{center}
The $\Z_2$ symmetry forbids
\begin{equation}
  (\wt QQ) \Sigma ~,
\end{equation}
on the other hand,
\begin{equation}
  \frac{1}{M^{10}}(\wt QQ)\Sigma (\det \wt QQ)
\end{equation}
is allowed
\footnote{
  By this change, a run away supersymmetric vacuum appears. However,
  we do not mind the vacuum because the large VEV which is much larger
  than Planck scale is required for making the potential energy
  smaller than at our local minimum point. 
  }.
Consequently the F-term of the $\Sigma$ becomes naturally small as
\begin{equation}
  \VEV{F_{\Sigma}}\sim \left( \frac {\Lambda}{M}\right)^{10} \Lambda^2.
\end{equation}
As written later by gauge messenger , SUSY breaking parameters in
visible sector
\begin{eqnarray}
  M_{SB}&\sim&  \frac { \alpha (\Lambda) }{4\pi}
  \frac {\VEV{F_{\Sigma}}}{\VEV{\Sigma}} \\
  &\sim& \frac {\alpha (\Lambda)}{4\pi}\left( 
    \frac{\Lambda}{M}\right)^{10} \Lambda \label{7}
\end{eqnarray}
are induced. Moreover, 
\begin{eqnarray}
  &&+\frac{1}{M^9}\det (\wt QQ) H_a \wt \phi \\
  &&+ \frac{1}{M^{11}}\det (\wt QQ)(\wt QQ)(a_{a}H_a\wt D_1 
  +b\phi \wt D_2+ c_a\phi \wt T_a) 
\end{eqnarray}
are allowed
\footnote{
Other allowed terms do not play physically important roles.
}, which induce the supersymmetric Higgs ($H_u$ and $\wt
H_d$) mass term
\begin{equation}
  \mu \sim \left( \frac {\Lambda}{M}\right)^{11}\Lambda ~.
\end{equation}
This can be a solution for the $\mu$ problem in this model when 
\begin{equation}
  \frac{\alpha (\Lambda)}{4\pi}\sim \frac{\Lambda}{M} ~.\label{CondOfMu}
\end{equation}
You should notice that this
solution for the $\mu$ problem is realized only 
in the scenario that SUSY and GUT gauge symmetry\ is broken
simultaneously. When we take the GUT scale $10^{15}~{\rm GeV}
\le \Lambda\le 10^{16}~{\rm GeV}$ and the supersymmetric Higgs mass 
$\mu\sim 100~{\rm GeV}$, we should take $\frac{\Lambda}{M}\sim 0.06$. 
Of course, we think that this value may be easily changed by one order
because of some ambiguities in the eqs. (\ref{3}) and (\ref{5})
\cite{nda}.
Therefore we do not remove the possibility that  $M$ is regarded as the
reduced Plank mass $M_{{\rm pl}}$ ($\sim 10^{18}~$GeV), though
it is difficult that the
gauge coupling does not diverge until the Plank scale in this model.

The Higgs mass mixing term (B-term) is often larger than the $\mu$
term in gauge mediated SUSY breaking models. But it is not the case
in this model,
because the origin of the $\mu$ term is not the same as the origin
of the SUSY breaking.
Actually, using the Spurion technique  developed
by Giudice and Rattazzi \cite{giudice}, we estimate the SUSY breaking
parameters
in the following case. The gauge group $G\times H$ is broken to the
$L$ gauge group by $\VEV{\Sigma}\sim \Lambda $ with the relation
$\alpha^{-1}_L(\Lambda)=\alpha_G^{-1}(\Lambda)+\alpha_H^{-1}(\Lambda)$
and the F component
$\VEV{F_\Sigma}$ is non zero.
Then the gaugino mass $M_i$, the scalar soft mass square $\tilde m^2$
of the matter field 
charged under $G$ and $L$ and the scalar two or three point vertex
$A_i$ are
\begin{eqnarray}
  M_a(\mu) &=& \frac{\alpha_L(\mu)}{4\pi}(b_L-b_G-b_H)
  \frac{\VEV{F_{\Sigma}}}{\VEV{\Sigma}}~,
  \\
  {\tilde m}^2(\mu) &=& 2C_G~ \frac{\alpha^2_L(\mu )}{(4\pi)^2}~
  \left\{ 
    \frac{\alpha^2_L(\Lambda)}{\alpha^2_L(\mu)}R^2b_G
    -\frac{\alpha^2_L(\Lambda)}{\alpha^2_L(\mu)}\frac{C_L}{C_G}
    \frac{(b_H+b_G)^2}{b_L}
  \right.\nn\\
  &&{}\left.
    +\frac{C_L}{C_G}\frac{(b_H+b_G-b_L)^2}{b_L}
  \right\}\left|\frac{\VEV{F_{\Sigma}}}{\VEV{\Sigma}}\right|^2 ~,
  \label{8}\\
  A_i(\mu)&=& 2C_G\frac{\alpha_L(\mu)}{4\pi}
  \left[R+\frac{C_L}{C_G}\frac{b_G+b_H-b_L}{b_L}
    -\frac{C_L}{C_G}\frac{b_G+b_H}{b_L}
    \frac{\alpha_L(\Lambda)}{\alpha_L(\mu)}
  \right]\frac{\VEV{F_{\Sigma}}}{\VEV{\Sigma}} ~,
  \\
  R &\equiv& \frac{\alpha_G(\Lambda)}{\alpha_L(\Lambda)}
  =1+\frac{\alpha_G(\Lambda)}{\alpha_H(\Lambda)}~,
\end{eqnarray}
where $b_H, b_G$ and $b_L$ are the coefficients of 1-loop $\beta$
functions of $H$, $G$ and $L$ around the scale $\Lambda$, respectively, 
and $C_G$ and $C_L$ are the quadratic Casimir of the field in $G$ and
$L$ ($C= (N^2-1)/2N$ for an $SU(N)$ fundamental field).
From the above expression and the definition of 
$A_i$ (${\cal L}=\sum_i A_i Q_i \del_{Q_i} W(Q)+h.c.$ where the $Q_i$
is the scalar component of the superfield $Q_i$), the Higgs mixing is
roughly estimated by $B\sim \mu M_{SB}$.

In the above expression the effect of Yukawa coupling is neglected. 
We must include the effect of top
Yukawa coupling and discuss on the renormalization group analysis more
carefully. We only comment that since the gauge theory is asymptotically
non free, various parameters are enhanced and there are some parameter 
region where all the scalar soft mass
squares except of the $H_u$ become positive. We will study the detail
in the next paper.

Finally, we should comment on the phenomenological constraints. 
The proton decay by dimension 5 operator \cite{dim5} is strongly
suppressed by the $\Z_2$ symmetry, though the proton decay by
dimension 6 operators may be seen.  Roughly speaking, the scale
$\Lambda$ should be larger than $10^{15}$ GeV. 
The gravitino mass $m_{3/2}\sim \VEV{F_{\Sigma}}/M_{{\rm pl}}$ 
should be smaller than 10 GeV for the phenomenological constraints
from the flavour changing neutral
currents ($K^0 \wt K^0$ mixing, $\mu \ra e \gamma$, etc.).  
If we take $\VEV{F_{\Sigma}}/\Lambda=10^{3}$, then $\Lambda \le
10^{16}$ GeV. In this model if there is
no CP phase in the VEV of the $\Sigma_Q$ and $\Sigma$ fields,
all CP violation phase except Kobayashi-Maskawa phase can be removed
by field redefinition.

In summary, we construct a model in which the SUSY and the GUT gauge
symmetry are dynamically broken simultaneously. The model is free for
the doublet-triplet splitting problem and the $\mu$ problem. 

We would like to thank K.Yoshioka for valuable comments on the 
effects of the renormalization group.
The work of T.H. is supported in part by the Grant-in-Aid
for JSPS fellows. The work of N.M. is supported in part by the
Grant-in-Aid for Scientific Research from the Ministry of Education,
Science and Culture.


\end{document}